# An Algorithm to Realize Rate Splitting Multiple Access in Gaussian Multi-Access Channel

Xiaomao Mao[*], Qinru Qiu[†], Huifang Chen[*], and Peiliang Qiu[*]

[*]Department of Information Science and Electronic Engineering,

Zhejiang University, Hangzhou, 310027, China

Email:{maoxm, chenhf, qiupl}@zju.edu.cn

[†]Department of Electrical and Computer Engineering,

Binghamton University, Binghamton, NY, 13902, USA

Email:qqiu@binghamton.edu

### Abstract

Rate Splitting Multiple Access (RSMA) is a code division multi-access technique which can achieve any base in the multi-access capacity polymatroid without high coding complexity or synchronization among the transmitting users. In this paper, a practical algorithm is proposed to compute the splitting coefficients and the successive decoding order of virtual users in Gaussian RSMA transmission. Based on the proposed algorithm, a deterministic mapping is built between the system parameters and the objective rate tuple for RSMA. As a result, the application of the RSMA technique becomes possible in current communication systems.

This work is supported by National Natural Science Foundation of China, Grant No.60772093.

## I. MOTIVATION

Denote the transmitted signal of user $i$, $i \in I = \{1, ..., N\}$, as $X_i$ and the received signal at the receiver as $Y$. The Gaussian multi-access channel can be modeled by

$$Y = \sum_{i=1}^{N} X_i + Z, \qquad (1)$$

where $Z$ denotes the Additive White Gaussian Noise (AWGN) and is distributed as $\mathcal{N}(0, \sigma^2)$. The capacity region of such a Gaussian multi-access channel possesses a polymatroid structure [5], [6].

For any $\mathbf{X} = (X_1, X_2, ..., X_N) \in \mathbb{R}_+^N$, let

$$\mathbf{X}(A) \triangleq \sum_{i \in A} X_i, \forall A \subseteq I. \qquad (2)$$

*Definition 1* Let set function $\rho : 2^I \to \mathbb{R}_+$ satisfy

1) $\rho(\emptyset) = 0$,
2) $\rho(A) \leq \rho(B), A \subseteq B \subseteq I$,
3) $\rho(A) + \rho(B) \geq \rho(A \cup B) + \rho(A \cap B)$.

In other words, $\rho$ is a monotone nondecreasing submodular function with $\rho(\emptyset) = 0$. Then, the polyhedron

$$(I, \rho) = \left\{ \mathbf{X} \in \mathbb{R}_+^N \,|\, \mathbf{X}(A) \leq \rho(A), \forall A \subseteq I \right\}, \qquad (3)$$

is called a polymatroid, where $I$ is called the ground set and $\rho$ the rank function.

A vector $\mathbf{X} \in \mathbb{R}_+^N$ is called an independent vector of the polymatroid $(I, \rho)$ if it is contained in the polyhedron represented by (3). For any $\mathbf{X}$ and $\mathbf{Y}$ in $\mathbb{R}_+^N$, let a partial order relation $\preceq$ be defined by

$$X \preceq Y \Leftrightarrow X_i \leq Y_i, \forall i \in I. \qquad (4)$$

A base is an independent vector which is maximal with respect to the partial order relation $\preceq$. The sum of elements is equal for all the bases. The set containing all the bases is called the dominant face. The dominant face is a convex hull of the extreme points of the polymatroid, and these extreme points are called the vertices.

Given the transmit power constraints $\mathbf{P} = (P_1, ..., P_N)$, the multi-access capacity polymatroid is

$$C_{MAC}(\mathbf{P}, \sigma^2) = \left\{ \mathbf{R} \in \mathbb{R}_+^N \,\bigg|\, \mathbf{R}(A) \leq \frac{1}{2} \log\left(1 + \frac{\mathbf{P}(A)}{\sigma^2}\right), A \subseteq I \right\}, \qquad (5)$$

where $\sigma^2$ is the Gaussian noise power. Essentially, the capacity polymatroid imposes $2^N$ constraints on a rate vector $\mathbf{R} = (R_1, ..., R_N)$ in the polymatroid.

Denote the dominant face of $C_{MAC}(\mathbf{P}, \sigma^2)$ as $\mathcal{K}_{MAC}(\mathbf{P}, \sigma^2)$. $\mathcal{K}_{MAC}(\mathbf{P}, \sigma^2)$ is the convex hull of the vertices $V_{\pi_1}, ..., V_{\pi_{N!}}$. Rate vectors at the vertices can be realized by successive decoding. More precisely, first one user is decoded, treating all other users as noise, then its decoded signal is subtracted from the received signal, then the next user is decoded and subtracted, and so forth. The rate vector at a vertex $V_{\pi_k}$ is determined by the successive decoding order $\pi_k$, which is a permutation on set $I$. There are $N!$ different permutations on set $I$, which correspond to $N!$ different successive decoding orders as well as to $N!$ different vertices. Varying the successive decoding order, we can achieve any vertex of the polymatroid. Denote the rate vector at a vertex $V_{\pi_k}$ as $(R_{\pi_k^{-1}(1)}, ..., R_{\pi_k^{-1}(N)})$ and its entries can be computed as

$$R_{\pi_k(i)} = \frac{1}{2} \log \left(1 + \frac{P_{\pi_k(i)}}{\sigma^2 + \sum_{l=1}^{i-1} P_{\pi_k(l)}}\right). \tag{6}$$

At the receiver, the corresponding successive decoding order is the inverse order of $\pi_k$, i.e., $\pi_k(N) \to \pi_k(N-1) \to ... \to \pi_k(1)$.

When $\mathbf{R} \in \mathcal{K}_{MAC}(\mathbf{P}, \sigma^2)$,

$$R_{sum} = \sum_{i=1}^{N} R_i = \frac{1}{2} \log \left(1 + \frac{\sum_{i=1}^{N} P_i}{\sigma^2}\right) \tag{7}$$

Namely, the sum rate of any base in the dominant face $\mathcal{K}_{MAC}(\mathbf{P}, \sigma^2)$ is equal to $R_{sum}$.

Because any rate vector in the capacity polymatroid $C_{MAC}(\mathbf{P}, \sigma^2)$ is dominated (with respect to the partial order relation) by a base in the dominant face, in this paper we investigate the methods to achieve a base in the dominant face $\mathcal{K}_{MAC}(\mathbf{P}, \sigma^2)$. Generally, there are three methods:

- Time sharing among the vertices;
- Joint encoding/decoding;
- Rate Splitting Multiple Access (RSMA).

Time sharing based method has high implementation complexity. To realize a base in the dominant face, the time sharing transmission requires as many as $N$ multi-access codes, each of which consists of $N$ single user codes, thus on the order of $N^2$ single user codes. Besides, synchronization is required among the users. The joint encoding/decoding based approach is also difficult to implement in practice because random codes have a decoding complexity of the order of $2^{mR_{sum}}$, where $m$ is the block length. Compared with time sharing among the vertices or joint encoding/decoding of the users, the RSMA technique significantly reduces the system implementation complexity. It requires at most $2N - 1$ single user codes to realize a base in the *N*-user multi-access capacity polymatroid, and synchronization is not required among the users.

In [3], RSMA scheme in the Gaussian multi-access channel is discussed and its feasibility is proved. Specifically, each user except one is first split into two virtual users, and the associated splitting coefficients is denoted as $\epsilon = (\epsilon_1, ..., \epsilon_N)$. For a user $i$, if $\epsilon_i(1 - \epsilon_i) \neq 0$, its power is split into two parts, $\epsilon_i P_i$ and $(1 - \epsilon_i)P_i$, each of which corresponds to a virtual user. If $\epsilon_i(1 - \epsilon_i) = 0$, it means that user $i$ is unsplit. In the next, a decoding order $\pi$ of the virtual users is found to realize the objective rate tuple by successive decoding. Although the proof in [3] is rigorous, no practical algorithm is proposed to compute the splitting parameters. In [1], the RSMA technique is extended to Discrete Memoryless Channel and an exhaustive searching algorithm is implied in the proof. In [2], a simple relation between system parameters and the resulting rate tuple is developed for RSMA, however, by using switches controlled by a sequence of independent and identically distributed (i.i.d) random variables. Due to high computation or implementation complexity, these two methods are both impractical to apply in current communication systems.

In order to realize a rate tuple in the dominant face, i.e. a base of the polymatroid, by RSMA, we propose an algorithm to compute the splitting coefficients $\epsilon$ as well as the corresponding decoding order $\pi$. In this way, the deterministic mapping is build between the splitting parameters and the resulting rate tuple in the Gaussian RSMA transmission. The rest of the paper is arranged as follow. In Section II, we introduce some definitions and extend some previously known results in [3] for Gaussian multi-access capacity region. In Section III, we propose the method to compute the splitting parameters for RSMA transmission. The proposed algorithm includes two parts. In the first part, the splitting order is determined from a combination process. Based on the obtained splitting order, we reduce the *N*-user rate splitting process to a series of 2-user rate splitting processes, and then propose the methods to compute the splitting parameters for such 2-user cases in the second part. We also give an example to illustrate the splitting procedure for the proposed algorithm. At last, we conclude the paper in Section IV.

## II. A Brief Review

Let $g(P, \sigma^2)$ denote the capacity of a single user AWGN channel with transmit power $P$ and noise power $\sigma^2$, i.e,

$$g(P, \sigma^2) = \frac{1}{2} \log\left(1 + \frac{P}{\sigma^2}\right). \tag{8}$$

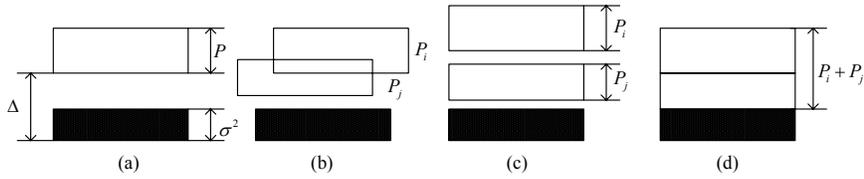

Fig. 1. Convenient representation: different RA types for a $N = 2$ multi-access transmission.

It can be easily verified that

$$g(P, \sigma^2) = g(P_1, \sigma^2) + g(P_2, P_1 + \sigma^2) \qquad (9)$$

is valid for all nonnegative numbers $P_1$, $P_2$ and $\sigma^2$ if $P_1 + P_2 = P$.

Let $f(R, P)$ denote the noise and interference power sum (NIS) $\Delta$ that a user with power $P$ can tolerate to realize a rate $R$ in a single user AWGN channel. Then,

$$f(R, P) = \sigma^2 + \delta = \frac{P}{2^{2R} - 1}, \qquad (10)$$

where $\delta$ denotes the power of other Gaussian interferences. Given the transmit power $P$, the rate $R$ determines the NIS $\Delta$ and vice versa.

Let $(N, \mathbf{P}, \mathbf{R}, \sigma^2)$ represent the feasible rate allocation (RA) in a $N$-user Gaussian multi-access capacity polymatroid with power constraint $\mathbf{P} = (P_1, ..., P_N)$ and noise power $\sigma^2$. For a RA, if (7) is satisfied, the RA is called *tight*. In other words, the RA of a base in the capacity polymatroid is a tight RA.

Adopting a similar method as reference [3], we associate a representation to each user in the multi-access transmission. As shown in Fig. 1 (a), the bottom black box represents the Gaussian noise and its height is proportional to the noise power $\sigma^2$. The transmitting user is represented by a white rectangle whose height and vertical position are proportional to its power $P$ and NIS $\Delta$, respectively. The nonzero widths of the rectangles and their horizontal positions are chosen for convenience and bear no information. Several possible scenarios for 2-user RAs are shown in Fig. 1 (b)-(d). Specifically, for user $i$ and user $j$,

- Case (b) denotes an overlapping RA.

$$\Delta_j \leq \Delta_i < \Delta_j + P_j. \qquad (11)$$

The corresponding rate tuple is not successive-decoding-achievable, but the RA could be tight, for example a non-vertex base in the dominant face.

- Case (c) denotes a discontiguous RA.

$$\Delta_i > \Delta_j + P_j. \tag{12}$$

The corresponding rate tuple is successive-decoding-achievable, but the RA is not tight, i.e., not a base.

- Case (d) denotes a contiguous RA.

$$\Delta_i = P_j + \Delta_j. \tag{13}$$

The RA is tight and nonoverlapping. The corresponding rate tuple is successive-decoding-achievable. In fact, it is a vertex in the capacity polymatroid.

A RSMA transmission is demanded for the overlapping RAs like the one in case (b). While, for those in case (c) and case (d), single user encoding combined with successive decoding at the receiver is enough to achieve the objective rate tuple.

*Lemma 1* For user $i$ and $j$, their transmit power tuple and rate tuple are $(P_i, P_j)$ and $(R_i, R_j)$, respectively. Let $\Delta_k = f(R_k, P_k)$ denote the NIS of user $k$, $k = i, j$. If the 2-user RA is not an overlapping RA, i.e.,

$$\Delta_i \geq \Delta_j + P_j, \tag{14}$$

then let $\Delta = f(R_i + R_j, P_i + P_j)$, we have

$$\Delta_j \leq \Delta \leq \Delta_i - P_j. \tag{15}$$

*Proof.* The first inequality in (15) follows from

$$g(P_i + P_j, \Delta) = R_i + R_j = g(P_i, \Delta_i) + g(P_j, \Delta_j)$$

$$\leq g(P_i, \Delta_j + P_j) + g(P_j, \Delta_j)$$

$$= g(P_i + P_j, \Delta_j).$$

Note that strict inequality in (14) implies strict inequality in (15). The second inequality in (15) follows from

$$g(P_i + P_j, \Delta) = R_i + R_j = g(P_i, \Delta_i) + g(P_j, \Delta_j)$$

$$\geq g(P_i, \Delta_i) + g(P_j, \Delta_i - P_j)$$

$$= g(P_i + P_j, \Delta_i - P_j).$$

□

**Remark 1** For a nonoverlapping RA, if the two "nonoverlapping" users are combined by putting their power together, in order to keep their sum rate constant, the lower user has to "move up" while the upper user has to "move down".

*Lemma 2* For user $i$ and $j$, their transmit power tuple and rate tuple are $(P_i, P_j)$ and $(R_i, R_j)$ respectively. Let $\Delta_k = f(R_k, P_k)$ denote the NIS of user $k$, $k = i, j$. If the 2-user RA is an overlapping RA, i.e.,

$$\Delta_j \leq \Delta_i < \Delta_j + P_j, \tag{16}$$

then let $\Delta = f(R_i + R_j, P_i + P_j)$, we have

$$\Delta_i - P_j \leq \Delta < \Delta_j. \tag{17}$$

*Proof.* The first inequality in (17) follows from

$$g(P_i + P_j, \Delta) = R_i + R_j = g(P_i, \Delta_i) + g(P_j, \Delta_j)$$

$$\leq g(P_i, \Delta_i) + g(P_j, \Delta_i - P_j)$$

$$= g(P_i + P_j, \Delta_i - P_j).$$

The second inequality in (17) follows from

$$g(P_i + P_j, \Delta) = R_i + R_j = g(P_i, \Delta_i) + g(P_j, \Delta_j)$$

$$\geq g(P_i, \Delta_j + P_j) + g(P_j, \Delta_j)$$

$$= g(P_i + P_j, \Delta_j).$$

$\square$

**Remark 2** We combine the two "overlapping" users in an overlapping RA into a superuser by putting their power together. If the superuser realizes the sum rate of the users, it has to "cover up" the upper user.

*Lemma 3* If an $N$-user RA is tight, then at least two among the $N$ users have an overlapping or contiguous RA.

*Proof.* Sort the users in an increasing order of their NISs and label them accordingly,

$$\sigma^2 \leq \Delta_1 \leq ... \leq \Delta_N.$$

Assume that there is no overlapping or contiguous RA among the users, i.e.,

$$\Delta_{i+1} > \Delta_i + P_i, \forall i = 1, ..., (N-1),$$

then we have

$$\sum_{i=1}^{N} R_i = \sum_{i=1}^{N} g(P_i, \Delta_i)$$

$$< \sum_{i=1}^{N} g(P_i, \Delta_1 + \sum_{k<i} P_k)$$

$$= g(\sum_{i=1}^{N} P_i, \Delta_1) \leq g(\sum_{i=1}^{N} P_i, \sigma^2),$$

which contradicts the assumption that the *N*-user RA is tight. Therefore, we arrive at the conclusion that in a tight *N*-user RA at least two among the users have an overlapping or contiguous RA. □

*Lemma 4* (Theorem I.1 in [4], as we paraphrase) If the rates of a subset of users lie on the dominant face of the polymatroid, then the subset of users may be replaced by a superuser that transmits at the combined rate with the combined power of the subset of the users.

**Remark 3** For a tight *N*-user RA, i.e. a base in the *N*-user polymatroid, there are two among the users have an overlapping or contiguous RA. If we combine these two users into a superuser transmitting at the sum rate with the sum power of the two users, and substitute the superuser for them in the user set, we obtain an $(N-1)$-user RA. The obtained $(N-1)$-user RA is also tight, i.e. a base in the $(N-1)$-user polymatroid.

## III. Compute the Splitting Parameters for RSMA transmission in Gaussian Multi-access Channel

With the power constraints for the users, the rate splitting approach converts an overlapping RA in the *N*-user multi-access capacity polymatroid to a nonoverlapping RA in the (2*N*-1)-user multi-access capacity polymatroid. As a result, single user encoding combined with successive decoding at the receiver is enough to achieve the objective rate allocation. In this section, we will propose a method to compute the splitting coefficients and the successive decoding order in the RSMA transmission. The proposed method is inspired by the proof of theorem 1 in [3]. However, it tackles the problem from a totally different aspect. We will first determine a splitting order based on which the *N*-user ($N \geq 3$) rate splitting process reduces to a series of 2-user (with overlapping or contiguous RA) rate splitting processes. Then, we compute the splitting coefficients $\epsilon$ as well as the successive decoding order $\pi$ for such 2-user rate splitting. By repeating the computation for the 2-user rate splitting according to the obtained splitting order, we obtain the overall *N*-user rate splitting parameters.

### A. Compute the Splitting Order

For a tight *N*-user RA $(N, \mathbf{P}, \mathbf{R}, \sigma^2)$, we can compute the NIS tuple $\Delta = (\Delta_1, ..., \Delta_N)$ from (10). Then, we sort the users in an increasing order of their NISs and relabel them accordingly. From Lemma 3, we know at least two among the *N* users have an overlapping or contiguous RA. Assume these two users are user $i$ and user $j$, and their power tuple and rate tuple are $(P_i, R_i)$ and $(P_j, R_j)$, respectively. We combine these two users into a superuser with transmit power $P_{i+j} = P_i + P_j$ and rate $R_{i+j} = R_i + R_j$. The NIS of the superuser is $f(P_i + P_j, R_i + R_j)$. Substituting this superuser for user $i$ and user $j$, we obtain an $(N-1)$-user RA $(N-1, \mathbf{P}', \mathbf{R}', \sigma^2)$, where $\mathbf{P}' = (P_1, ..., P_{i+j}, ..., P_N)$ and $\mathbf{R}' = (R_1, ..., R_{i+j}, ..., R_N)$. Albeit with $N-1$ users, the obtained $(N-1)$-user RA is tight, i.e. a base in the $(N-1)$-user multi-access polymatroid. Therefore, at least two among the users have an overlapping or contiguous RA. Similarly, we can find these two users and combine them into a superuser, then substitute the superuser for them and obtain a tight $(N-2)$-user RA. We combine two users (or superusers) at a time and repeat the combination until we get one superuser with power $\sum_{i=1}^{N} P_i$ and rate $\sum_{i=1}^{N} R_i$. As shown in Fig. 2, the combination process can be described by a binary tree. The leaf nodes represent the users while the root node represents the last obtained superuser. From this binary tree, we can determine the splitting order which is just the inverse order of the combination process. In other words, the splitting process begins with the root node, and each split corresponds to a parent node being split into its two siblings. Note that the above mentioned combination process is feasible for any tight $N$-user RA. However, multiple combination orders, and hence multiple splitting orders, may exist for a given tight $N$-user RA.

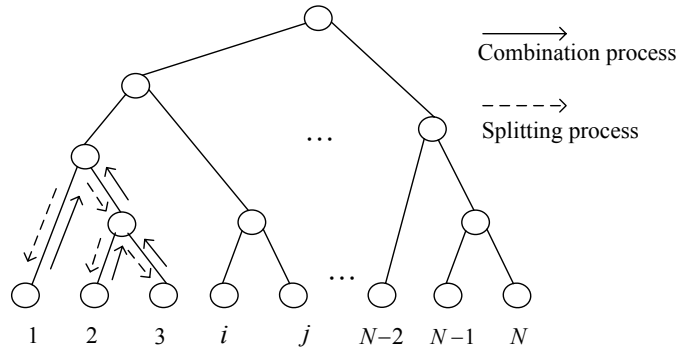

Fig. 2. The combination process can be described by a binary tree. From the combination binary tree, the splitting order can be determined.

*B. Compute the Splitting Coefficients and the Successive Decoding Order*

On obtaining the splitting order, the $N$-user rate splitting reduces to a series of 2-user rate splitting. Intuitively, for a 2-user case, the rate splitting technique divides the white rectangle associated with each user in Fig. 1 case (b) into two rectangles, and then fills the white area in Fig. 1 case (d) using these rectangles without overlap. During this procedure, the size of these rectangles and their vertical positions are adjusted, such that not only the users' transmit powers but also their rates remain the same. As a result, successive decoding is enough to achieve the objective rate tuple.

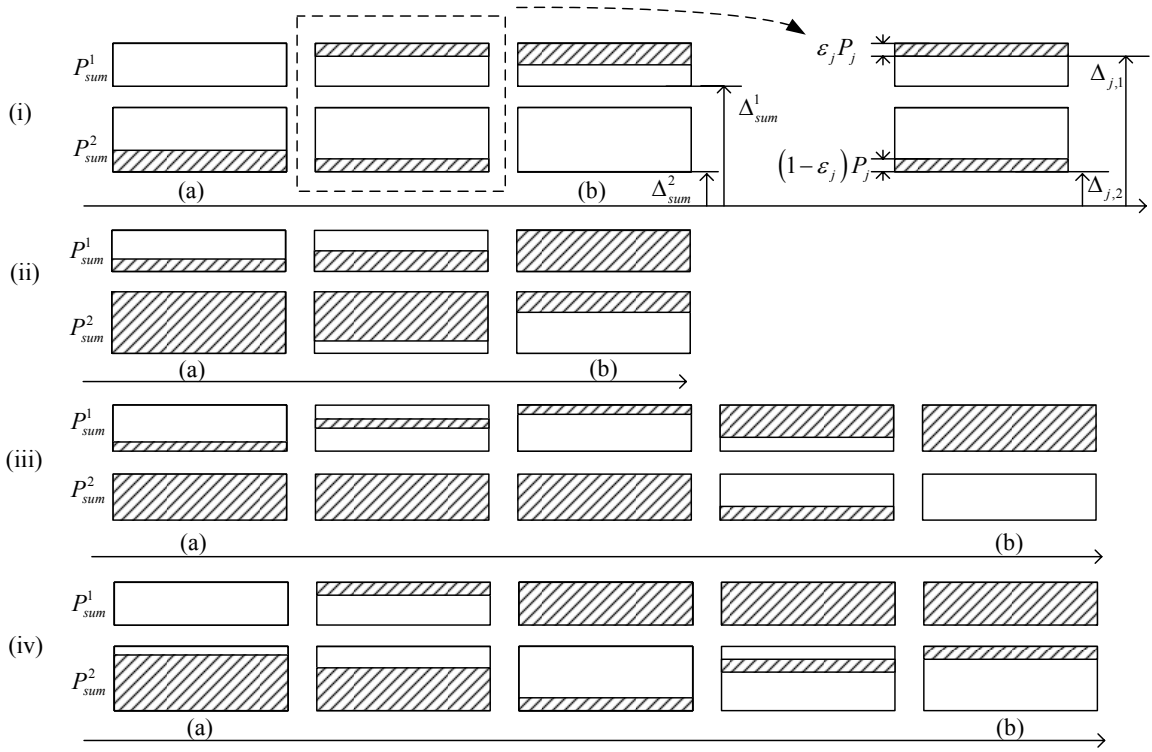

Fig. 3. The nonoverlapping power filling for different possible filling patterns.

Since the users do not have to split for the contiguous RAs, we only consider the overlapping RAs in the following analysis.

*Theorem 1* Assume that a 2-user RA with power tuple $(P_i, P_j)$ and rate tuple $(R_i, R_j)$ is an overlapping RA, i.e.,

$$\Delta_j \leq \Delta_i < \Delta_j + P_j. \tag{18}$$

Given another discontiguous 2-user RA with power tuple $(P_{sum}^1, P_{sum}^2)$ and rate tuple $(R_{sum}^1, R_{sum}^2)$,

$$\Delta_{sum}^1 > P_{sum}^2 + \Delta_{sum}^2, \tag{19}$$

if

$$\begin{aligned} P_{sum}^1 + P_{sum}^2 &= P_i + P_j, \\ R_{sum}^1 + R_{sum}^2 &= R_i + R_j, \end{aligned} \tag{20}$$

then let user $k$, $k = i, j$, be split associated with a coefficient $\epsilon_k$ into two virtual users with powers $P_{k,1}$ and $P_{k,2}$,

$$P_{k,1} = \epsilon_k P_k, P_{k,2} = (1 - \epsilon_k) P_k, k = i, j, \tag{21}$$

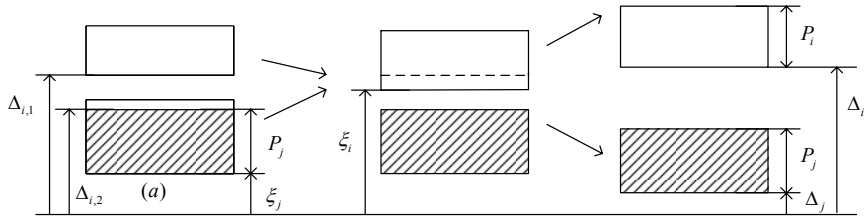

Fig. 4. Proof, by contradiction, that in pattern (a) $\rho_j \geq R_j$.

we obtain a 4-user RA with power tuple $(P_{i,1}, P_{i,2}, P_{j,1}, P_{j,2})$. Assume the corresponding rate tuple is $(R_{i,1}, R_{i,2}, R_{j,1}, R_{j,2})$. By adjusting the splitting coefficients $\epsilon_i$ and $\epsilon_j$ and the successive decoding order of the virtual users(i.e., the vertical positions of the virtual users), we can make the 4-user RA nonoverlapping. In addition, the sum rate $\rho_k$ of the virtual users that belong to user $k$, $k = i, j$, equals to $R_k$, i.e.,

$$\rho_i = R_{i,1} + R_{i,2} = R_i,$$
$$\rho_j = R_{j,1} + R_{j,2} = R_j. \quad (22)$$

*Proof.* In order to prove Theorem 1, we have to nonoverlappingly fill the rectangles of $P^1_{sum}$ and $P^2_{sum}$ with the virtual user's rectangles and find a filling pattern which realizes the target rate tuple $(R_i, R_j)$ by adjusting the sizes and the vertical positions of the virtual user's rectangles. The possible filling patterns are shown in Fig. 3 for case (i), $P^1_{sum} \geq P_j$, $P^2_{sum} \geq P_j$; case (ii), $P^1_{sum} \leq P_j$, $P^2_{sum} \leq P_j$; case (iii), $P^1_{sum} \geq P_j$, $P^2_{sum} \leq P_j$; case (iv), $P^1_{sum} \leq P_j$, $P^2_{sum} \geq P_j$. The white rectangles denote the virtual users that belong to user $i$, while the grey rectangles denote that of user $j$. If two virtual users that belong to the same user are contiguous in the filling, the user is considered unsplit. For each of the four cases, when going from pattern (a) to pattern (b), the sum rate $\rho_i$ of user $i$ increases continuously from its minimum to its maximum value. Since $\rho_i + \rho_j = R^1_{sum} + R^2_{sum} = R_i + R_j$ across all patterns, the sum rate $\rho_j$ of user $j$ accordingly decreases continuously from its maximum to its minimum. Here, a filling pattern specifies a decoding order of the virtual users as well as the corresponding splitting coefficients.

To prove the Theorem, we argue that in Fig. 3, $\rho_j \geq R_j$ for pattern (a) and $\rho_j \leq R_j$ for pattern (b). For convenience, pattern (a) in Fig. 3 case (iv) is redrawn in Fig. 4. Assume that the claim is false, i.e. for pattern (a) $\rho_j \leq R_j$, thus $\rho_i \geq R_i$. That is

$$\xi_j > \Delta_j, \xi_i < \Delta_i, \quad (23)$$

where, $\xi_k = f(\rho_k, P_k)$, $k = i, j$. As shown in Fig. 4, combining those two nonoverlapping virtual users of the user $i$, we have, from Lemma 1,

$$\xi_i \geq \xi_j + P_j. \tag{24}$$

Therefore, using (23) and (24), we have

$$\Delta_i > \xi_i \geq \xi_j + P_j > \Delta_j + P_j, \tag{25}$$

which contradicts the assumption in (18) which states that $\Delta_i < \Delta_j + P_j$. Similarly, we can prove that $\rho_j \leq R_j$ in pattern (b) of Fig. 3 case (iv) is necessary for the assumption $\Delta_j \leq \Delta_i$. Since the evolution from pattern (a) to pattern (b) is continuous, there must be some pattern that satisfies

$$\rho_i = R_{i,1} + R_{i,2} = R_i,$$

$$\rho_j = R_{j,1} + R_{j,2} = R_j.$$

For other cases in Fig. 3, the claim can be proved in the similar way.

What has been proved above indicates that the splitting coefficients as well as the successive decoding order of the virtual users can be determined according to the filling patterns in Fig. 3. Specifically, we can compute the splitting coefficient $\epsilon_j$ and the NISs of the virtual users for user $j$ as below.

Case (i):

$$\begin{aligned}
\epsilon_j &= \frac{(P_{sum}^1 + \Delta_{sum}^1)(\Delta_{sum}^1 \cdot 2^{2R_j} - P_j - \Delta_{sum}^2)}{P_j(\Delta_{sum}^2 \cdot 2^{2R_j} - P_{sum}^1 - \Delta_{sum}^1)}, \\
\Delta_{j,1} &= P_{sum}^1 + \Delta_{sum}^1 - \epsilon_j P_j, \\
\Delta_{j,2} &= \Delta_{sum}^2.
\end{aligned} \tag{26}$$

Case (ii):

$$\begin{aligned}
\epsilon_j &= 1 - \frac{(P_{sum}^2 + \Delta_{sum}^2)(\Delta_{sum}^1 \cdot 2^{2R_j} - P_j - \Delta_{sum}^1)}{P_j(\Delta_{sum}^1 \cdot 2^{2R_j} - P_{sum}^2 - \Delta_{sum}^2)}, \\
\Delta_{j,1} &= \Delta_{sum}^1, \\
\Delta_{j,2} &= P_{sum}^2 + \Delta_{sum}^2 - (1 - \epsilon)_j P_j.
\end{aligned} \tag{27}$$

Case (iii): If $R_j \geq g(P_{sum}^2, \Delta_{sum}^2) + g(P_j - P_{sum}^2, P_{sum}^1 + \Delta_{sum}^1 - (P_j - P_{sum}^2))$, then

$$\begin{aligned}
\epsilon_j &= 1 - \frac{P_{sum}^2}{P_j}, \\
\Delta_{j,1} &= f(R_j - g((1 - \epsilon_j)P_j, \Delta_{sum}^2), \epsilon_j P_j), \\
\Delta_{j,2} &= \Delta_{sum}^2.
\end{aligned} \tag{28}$$

If $R_j \leq g(P_{sum}^2, \Delta_{sum}^2) + g(P_j - P_{sum}^2, P_{sum}^1 + \Delta_{sum}^1 - (P_j - P_{sum}^2))$, then

$$\epsilon_j = \frac{(P_{sum}^1 + \Delta_{sum}^1)(\Delta_{sum}^1 \cdot 2^{2R_j} - P_j - \Delta_{sum}^2)}{P_j(\Delta_{sum}^2 \cdot 2^{2R_j} - P_{sum}^1 - \Delta_{sum}^1)},$$

$$\Delta_{j,1} = P_{sum}^1 + \Delta_{sum}^1 - \epsilon_j P_j, \tag{29}$$

$$\Delta_{j,2} = \Delta_{sum}^2.$$

Case (iv): If $R_j \geq g(P_{sum}^1, \Delta_{sum}^1) + g(P_j - P_{sum}^1, \Delta_{sum}^2)$, then

$$\epsilon_j = \frac{(P_{sum}^1 + \Delta_{sum}^1)(\Delta_{sum}^1 \cdot 2^{2R_j} - P_j - \Delta_{sum}^2)}{P_j(\Delta_{sum}^2 \cdot 2^{2R_j} - P_{sum}^1 - \Delta_{sum}^1)},$$

$$\Delta_{j,1} = P_{sum}^1 + \Delta_{sum}^1 - \epsilon_j P_j, \tag{30}$$

$$\Delta_{j,2} = \Delta_{sum}^2.$$

If $R_j \leq g(P_{sum}^1, \Delta_{sum}^1) + g(P_j - P_{sum}^1, \Delta_{sum}^2)$, then

$$\epsilon_j = \frac{P_{sum}^1}{P_j},$$

$$\Delta_{j,1} = \Delta_{sum}^1, \tag{31}$$

$$\Delta_{j,2} = f(R_j - g(\epsilon_j P_j, \Delta_{sum}^1), (1 - \epsilon_j)P_j).$$

Note that as long as the splitting parameters for user $j$ are determined, those for user $i$ can be computed correspondingly. □

If in Theorem 1 the 2-user RA with power tuple $(P_{sum}^1, P_{sum}^2)$ and rate tuple $(R_{sum}^1, R_{sum}^2)$ is a contiguous RA, i.e.,

$$\Delta_{sum}^1 = \Delta_{sum}^2 + P_{sum}^2,$$

only one of the users needs to be split. This case is described in the following Theorem.

*Theorem 2* Assume that a 2-user RA with power tuple $(P_i, P_j)$ and rate tuple $(R_i, R_j)$ is an overlapping RA, i.e.,

$$\Delta_j \leq \Delta_i < \Delta_j + P_j. \tag{32}$$

Given another 1-user RA with power $P_{sum}$ and rate $R_{sum}$, if

$$\begin{aligned} P_{sum} &= P_i + P_j, \\ R_{sum} &= R_i + R_j, \end{aligned} \tag{33}$$

then let one of the users, assume user $i$, remain unsplit while the other user, i.e. user $j$, be split into two virtual users associated with a splitting coefficient $\epsilon_j$,

$$P_{j,1} = \epsilon_j P_j,$$

$$P_{j,2} = (1 - \epsilon_j)P_j,$$

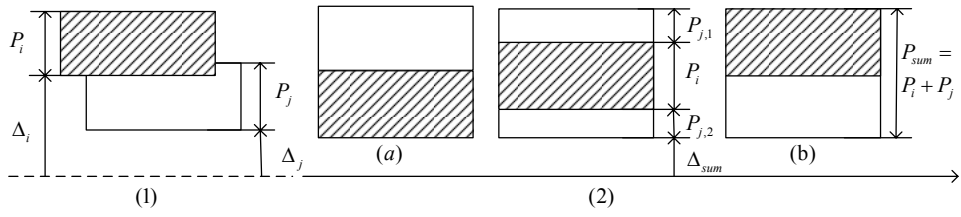

Fig. 5. The nonoverlapping power filling with one user unsplit.

we obtain a 3-user RA with power tuple $(P_{j,1}, P_{j,2}, P_i)$. Assume the corresponding rate tuple is $(R_{j,1}, R_{j,2}, R'_i)$. By adjusting the splitting coefficients $\epsilon_j$ and the successive decoding order of the virtual users (i.e the vertical positions of the virtual users), we can make the 3-user RA nonoverlapping. In addition, we can make

$$R_{j,1} + R_{j,2} = R_j$$
$$R'_i = R_i \qquad (34)$$

*Proof.* We prove this theorem in the similar way as in Theorem 1. In Fig. 5 case (1), it shows the overlapping 2-user RA, where the white rectangle represents user $j$ while the grey rectangle represents user $i$. As shown in Fig. 5 case (2), we fill the grey rectangle of user $i$ into the rectangle of $P_{sum}$. From pattern (a) to (b), user $i$'s rate $R'_i$ increases from its minimum to its maximum. Assume that $\Delta_{sum} = f(P_{sum}, R_{sum}) = f(P_i + P_j, R_i + R_j)$ and $\Delta_i = f(P_i, R_i)$. From Lemma 2, we have $\Delta_{sum} \leq \Delta_i$ and $P_{sum} + \Delta_{sum} \geq P_i + \Delta_i$. Therefore, there must be a pattern in which user $i$ realizes its rate, i.e $R'_i = R_i$. With the grey rectangle of user $i$ fixed, two white rectangles are resulted and they represent the virtual users of user $j$. It follows from the assumption and (9) that the sum rate of these virtual users achieves the rate $R_j$ of user $j$.

Specifically, we can compute the splitting parameters for user $j$ as below.

$$\epsilon_j = \frac{P_{sum} + \Delta_{sum} - (P_i + \Delta_i)}{P_j},$$
$$\Delta_{j,1} = P_i + \Delta_i, \qquad (35)$$
$$\Delta_{j,2} = \Delta_{sum}.$$

Since user $i$ is unsplit, the splitting coefficient $\epsilon_i = 1$ and the NIS $\Delta_i$ remains the same. □

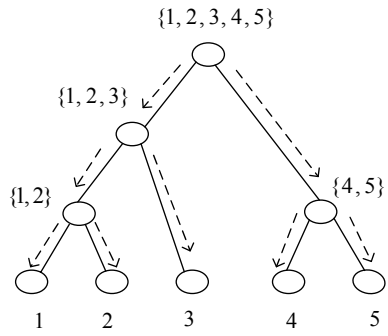

Fig. 6. The combination binary tree of a tight 5-user RA. The splitting is performed in the direction of the arrows.

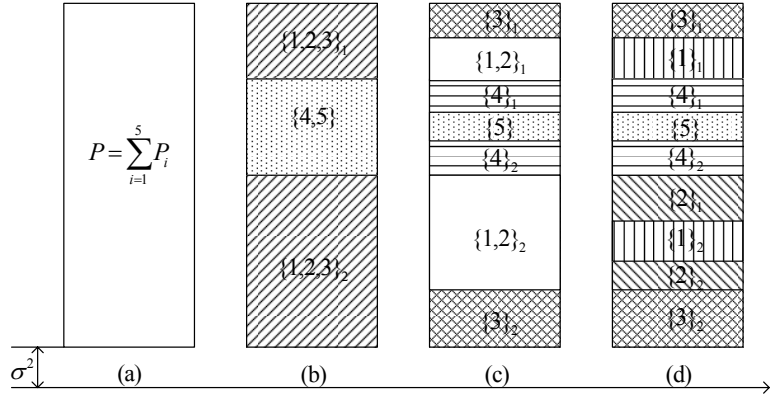

Fig. 7. The splitting process of a tight 5-user RA.

## C. The Rate Splitting Procedure for Gaussian Multi-access Transmission

For a tight RA $(N, \mathbf{P}, \mathbf{R}, \sigma^2)$, we first determine the splitting order from the acquired combination binary tree. One of such binary trees is shown in Fig. 2. Based on the splitting order, the *N*-user rate splitting reduces to a series of 2-user rate splitting. We compute the splitting parameters for such 2-user rate splitting by alternatively applying Theorem 1 and Theorem 2 for the overlapping RAs. As mentioned before, the users (or superusers) do not have to split for contiguous RAs. During the 2-user splitting process, we also record the NISs of the virtual users. Following the splitting order, we repeat the 2-user rate splitting until we acquire the splitting coefficients and the NISs for all leaf nodes in the combination binary tree, i.e the splitting parameters for all users. Sort the virtual users in an increasing order $\pi$ of their NISs, then we obtain the corresponding successive decoding order which is just the inverse order of $\pi$.

An example is shown in Fig. 6 and Fig. 7 for a 5-user rate splitting process. In Fig. 6,

the combination binary tree of the tight 5-user RA is shown. Following the splitting order which is determined from the combination binary tree, the splitting process evolves from pattern (a) to pattern (d) in Fig. 7. The splitting coefficients are computed from (17-22,26). The successive decoding is performed following the top-down order in Fig. 7 pattern (d). The first split in the splitting process either is for a contiguous case or fit in with Theorem 2. Therefore, at least one of the users (or superusers) involved in the first split remains unsplit. As a result, with the proposed method, the number of the resulted virtual users is at most $2N-1$ for a tight $N$-user RA. That corresponds to the complexity bound obtained in [3]. For the example shown in Fig. 7, the number of the virtual users is 9.

## IV. Conclusion

To achieve an objective rate tuple by the RSMA technique, we propose a practical algorithm to compute the splitting coefficients as well as the corresponding successive decoding order of the virtual users. The proposed algorithm includes two parts. A splitting order is first determined from the combination binary tree. Based on the splitting order, the $N$-user rate splitting reduces to a series of 2-user rate splitting. Then, we compute the splitting parameters for such 2-user cases. By repeating the 2-user rate splitting according to the splitting order, we obtain the overall splitting parameters for the $N$-user rate splitting. Though the proposed algorithm is analyzed for Gaussian RSMA transmission, it can be extended to solve the splitting parameters for RSMA in Discrete Memoryless Multi-access Channel. Based on the proposed algorithm, a deterministic mapping is built between the system parameters and the resulting rate tuple in RSMA transmission. As a result, the application of the $N$-user RSMA technique in current communication systems becomes possible.